\documentclass[twocolumn,showpacs,amsmath,amssymb,pra]{revtex4}

\usepackage{graphicx}%
\usepackage{dcolumn}
\usepackage{amsmath}
\newcommand{\be}{\begin{equation}}
\newcommand{\ee}{\end{equation}}
\newcommand{\bey}{\begin{eqnarray}}
\newcommand{\eey}{\end{eqnarray}}
\newcommand{\bw}{\begin{widetext}}
\newcommand{\ew}{\end{widetext}}

\newcommand{\ra}{\rangle}
\newcommand{\la}{\langle}

\newcommand{\E}{{\cal E}}

\begin{document}
 %\draft

 \title {Efficiency of dynamical decoupling sequences in presence of pulse errors
 }

 \author{Zhihao Xiao (肖之昊), Lewei He (何乐为), and Wen-ge Wang (王文阁)
 \footnote{ Email address: wgwang@ustc.edu.cn}}

 %\address{
 \affiliation{
 Department of Modern Physics, University of Science and Technology of China, 230026 China
 }

 \date{\today}

 \begin{abstract}

 For a generic dynamical decoupling sequence employing a single-axis control, 
 we study its efficiency in the presence of small errors in the direction of the controlling pulses. 
 In the case that the corresponding ideal dynamical-decoupling sequence produces
 sufficiently good results,
 the impact of the errors is found to scale as $\xi^2$, with negligible first-order effect,
 where $\xi$ is the dispersion of the random errors.
 This analytical prediction is numerically tested in a model, in which the environment is modeled by
 one qubit coupled to a quantum kicked rotator in chaotic motion. 
 In this model, with periodic pulses applied to the qubit in the environment,  
 it is found numerically that periodic bang-bang  control may outperform Uhrig dynamical decoupling.

 \end{abstract}


\maketitle

 %\begin{multicols}{2}

\section{Introduction}

 Dynamical decoupling (DD) has attracted lots of attention in the past years,
 due to its potential application in quantum information processes.
 The basic idea is to insert a sequence of controlling pulses within a time period of interest,
 such that the system of interest can be effectively decoupled from its environment.
 As a result, environment-induced decoherence can be effectively suppressed
 and initial coherence in the system can be preserved.

 Several DD schemes have been proposed.
 For example, the so-called periodic ``bang-bang '' control \cite{VL98,VKL99,Morton06}
 can suppress decoherence up to the order of $O(T^2)$ for a given period $T$
 of coherence preservation, 
 while the Carr-Purcell-Meiboom-Gill sequence has an $O(T^3)$ efficiency \cite{Slich,Uhrig07}. 
 A better result can be obtained by an approach called ``concatenated dynamical decoupling''
 \cite{KL05,KL07}, with an efficiency of the order of $O(T^{N+1})$ achieved by $2^N$ pulses.
 Recently, a remarkable progress has been made by Uhrig, showing that decoherence of a single
 spin, which is induced by a spin-boson bath,
 can be suppressed up to the order of $O(T^{N+1})$ with only $N$ pulses \cite{Uhrig07}.
 Later, the Uhrig dynamical decoupling (UDD) was conjectured \cite{LWS08} and rigorously proved to be
 model-independent for the pure dephasing case \cite{YL08}.
 More recently, it has been found that different types of UDD
 sequences may be integrated to obtain better results \cite{West2010}
 and,  with appropriate extension, to work for a system with two spins 
 as well \cite{MSSG10}.
 Meanwhile, the efficiencies of UDD and its generalization has been beautifully demonstrated 
 experimentally \cite{Bn09,Du09,Du11}

 In practical application of a DD sequence, an important topic is its robustness,
 i.e., the influence of non-idealness of the controlling pulses in the efficiency of the DD.
 The non-idealness may come from finite widths of the pulses and/or small deviation of the actual
 directions of the pulses from their designed directions.
 In the case that the non-ideal parts of the pulses with finite width
 satisfy certain symmetry requirements,
 a generalized UDD can be found \cite{YL08}.
 However, in a generic circumstance, accumulation effect of small imperfections in the pulses
 may have significant consequences when the number of pulses is not small.
 There have been several investigations in the impact of systematic pulse errors
 for some specific DD sequences, both experimentally \cite{AAPS10,Biercuk09,Wang10,Tyryshkin}
 and theoretically \cite{Wang10,Tyryshkin,WD11,KHCM10}.

 In this work, for a generic DD employing a single-axis control,
 we study accumulation effects of small random errors in the direction of the controlling pulses.
 Analytical derivations are given in Sec.\ref{sect-ana}. 
 In particular, we show that a DD, which has a sufficiently good performance in the ideal case,
 is robust in the sense that the pulse errors have negligible first-order effect.

 In numerical investigation, we employ a model in which there are periodic
 pluses in the environment, which are not directly applied to the qubit of interest.
 Using this model, we can test our analytical predictions, 
 and also study another topic of interest, namely, the influence of 
 high-frequency cutoff in the efficiency of UDD. 
 When the spectrum of the environment has a sharp high-frequency cutoff, 
 UDD has been found outperforming all other known DD sequences and is regarded as optimal.
 However, for environments with soft cutoffs in the spectra, there is no reason to expect that
 UDD is optimal; in fact, recently it has been shown that protocols with 
 periodic structure, such as the Carr-Purcell-Meiboom-Gill sequence, may have a better
 performance for this type of environment \cite{CLND08,PU10,DWRDH1011,AAS11}.

 Specifically,  In Sec.\ref{sect-num}, we study a model, in which the environment 
 is simulated by a second qubit coupled to a quantum kicked rotator in chaotic motion.
% The first qubit, whose coherence is to be preserved, is coupled to the qubit in the environment, 
% but not to the kicked rotator. 
 Previous study shows that this model, though a single-particle dynamical system,
 may simulate a pure-dephasing many-body bath
 (Caldeira-Leggett model \cite{CL83}), as well as some non-Markovian environments \cite{RCB06}.
 Our numerical simulations show that for this type of environment
 the periodic bang-bang  control may outperform UDD.
 Finally, conclusions and discussions are given in Sec.\ref{sect-conc}

\section{Analytical study of impacts of small pulse errors}
 \label{sect-ana}

 In this section, we first recall essential properties of DD, 
 then, derive expressions for the influence of small pulse errors in the efficiency of 
 a DD that employs a single-axis control.

 \subsection{Dynamical decoupling scheme}

 Let us first recall some essential properties of a generic DD.
 Consider a qubit $S$ and its environment $\E$, under a Hamiltonian
 \be H= H_S + H_I+H_{\E} , \label{Hi} \ee
 where $H_S$ and $H_\E$ are the Hamiltonians of $S$ and $\E$, respectively,
 and $H_I$ is the interaction Hamiltonian.
 The self-Hamiltonian $H_S$ is assumed to be a constant, which can be taken as zero, $H_S=0$.
 We further assume that the interaction Hamiltonian $H_I$ is commutable with $\sigma_z$,
 $[\sigma_z,H_I]=0$, where $\sigma_z$ is the $z$-component of the Pauli operator for the qubit $S$.

 Suppose the qubit $S$ lies initially in a state, 
 which is not an eigenstate of the interaction Hamiltonian $H_{I}$.
 Schr\"{o}dinger evolution under the Hamiltonian $H$, given by a unitary operator
 $U(T,0)$ for a time period $T$, may induce decoherence to the state of the qubit $S$.
 The purpose of a DD is to preserve coherence in the
 initial state of $S$ within the time period $T$ by inserting a sequence of pulses,
 e.g., $\pi$-pulses for the $x$ direction.
 Using $t_i=\delta_i T$ of $i=1, 2, \cdots, n$ to indicate the instants at which pulses are applied,
 with totally $n$ pulses, the time-evolution operator is now written as
 \bey R=U(T,t_n) \sigma_x U(t_n, t_{n-1})\sigma_x
   \cdots U(t_2,t_1)\sigma_xU(t_1,0). \label{R0}
 \eey

 To measure preservation of coherence in an initial state in the $x$-direction,
 one may consider measurement on the observable $\sigma_x$, which gives the signal
 \bey
 s(T)=\la\uparrow|{D^\dagger_y(-\pi/2)} {R^\dagger}{\sigma_x}R{D_y(-\pi/2)}|\uparrow\ra ,
 \eey
 where $|\uparrow\ra$ indicates an eigenstate of $\sigma_z$
 and $D_y(-\pi/2)$ rotates it to an eigenstate of $\sigma_x$.
 After some derivation, one gets
 \bey
 s(T)=\mathfrak{Re}\la\downarrow|{R^\dagger}{\sigma_x}R|\uparrow\ra . \label{sT}
 \eey
 The coherence is perfectly preserved if $s(T)=1$.

 In UDD, $\pi$-pulses are applied to the qubit $S$ at times $t_j={\delta_j}T$,
 where
 \bey
 \delta_j=\sin^2[{\pi}j/(2n+2)]\ \ \ \ \ (j=0,1,2,{\cdot\cdot\cdot}n,n+1) .
 \eey
 For UDD, $s(T)=1-{O}({T}^{2n+2})$.
 In a periodic bang-bang  control of DD, pulses are applied at times with $\delta_j = j/n$.

 \subsection{$s(T)$ expanded to the second-order term of error}
 \label{sect-deri}

 As discussed in Introduction, controlling pulses in a DD may be subject to small random errors 
 in its direction. 
 Let us consider small random deviation in the $y$ direction for $\pi$-pulses in the $x$ direction.
 In this case, $\sigma_x$ in the ideal time-evolution operator in Eq.~(\ref{R0}) should be 
 replaced by
 \be \sigma_\varepsilon = \varepsilon_x \sigma_x+\varepsilon_y\sigma_y , \label{sig-e}
 \ee
 where $\varepsilon_x =\sqrt{1-\varepsilon_y^2}$ and
 $\varepsilon_y$ is a small random number with Gaussian distribution,
 \bey f(\varepsilon_y)=\frac{1}{\sqrt{2\pi\xi^2}}\exp \left( {-\frac{\varepsilon_y^2}{2\xi^2}}
 \right ).
% \\ f(\varepsilon_z)=\frac{1}{\sqrt{2\pi\xi^2}}\exp \left( {-\frac{\varepsilon_z^2}{2\xi^2}}
% \right ).
 \eey
 Here, $\xi$ is the dispersion of the random number, with $\xi \ll 1$.
 Then, the time evolution operator, denoted by $R_\varepsilon$, is written as 
\bey
 R_{\varepsilon}&=&U(T,t_n)(\sqrt{1-\varepsilon_{y,n}^2}\sigma_x
 +\varepsilon_{y,n}\sigma_y)  \nonumber
 \\
 & & \cdot U(t_n,t_{n-1})(\sqrt{1-\varepsilon_{y,n-1}^2}\sigma_x+\varepsilon_{y,n-1}\sigma_y)
  \nonumber
 \\  &&\cdots U(t_2,t_1)(\sqrt{1-\varepsilon_{y,1}^2}\sigma_x+\varepsilon_{y,1}
 \sigma_y)U(t_1,0), \label{Re}
 \eey
 where we use $\varepsilon_{y,i}$ to indicate the value of $\varepsilon_y$ for the $i$-th pulse.
 Now, the signal, denoted by $s_\varepsilon(T)$, has the following expression,
 \bey
 s_\varepsilon(T) = \mathfrak{Re}\la\downarrow|{R_\varepsilon^\dagger}{\sigma_x}
 R_\varepsilon|\uparrow\ra . \label{st}
 \eey

 To evaluate $s_\varepsilon(T)$, we substitute Eq.~(\ref{Re}) into Eq.~(\ref{st}) and 
 expand the result in the power of $\varepsilon_y $.
 Up to the second-order term of $\xi$, we write the signal in the following form, 
 \bey
 s_\varepsilon(T)  = s_{\varepsilon0}(T)+s_{\varepsilon1}(T)+s_{\varepsilon2}(T)+
 O(\xi^3). \  
 \label{se} \eey
 Definitions of the first three terms on the right hand side of Eq.~(\ref{se}) will be 
 given below, when they are treated separately.

 The first term on the right hand side of Eq.~(\ref{se}) is obtained by considering
 only the contribution of $\varepsilon_x \sigma_x$ in each $\sigma_\varepsilon$, 
 \bey
 s_{\varepsilon0}(T)=\mathfrak{Re}\la\downarrow|{R_{\varepsilon0}^\dagger}
 {\sigma_x}R_{\varepsilon0}|\uparrow\ra , \label{s0-i}
 \eey
 where 
 \bey
  R_{\varepsilon0} = U(T,t_n)\sqrt{1-\varepsilon_{y,n}^2}
  \sigma_xU(t_n,t_{n-1}) \hspace{2cm}
  \nonumber\\
  \cdot \sqrt{1-\varepsilon_{y,n-1}^2}\sigma_xU(t_{n-1},t_{n-2}) \cdots
%  U(t_2,t_1)
%  \nonumber\\
  \sqrt{1-\varepsilon_{y,1}^2}\sigma_xU(t_1,0)
  \nonumber\\
  =\prod_{k=1}^n(1-\varepsilon_{y,k}^2)^\frac{1}{2}R. \hspace{0.3cm}
 \label{Re0} \eey
 Here, $R$ is the time evolution operator for the case of ideal pulses in Eq.~(\ref{R0}).
 Substituting Eq.~(\ref{Re0}) into Eq.~(\ref{s0-i}), we find
 \be s_{\varepsilon0}(T)=\prod_{k=1}^n(1-\varepsilon_{y,k}^2)s(T).  \ee
 It is seen that, for sufficiently small $\xi$,
 \be s_{\varepsilon0}(T) -s(T) \sim - n\xi^2 s(T). \label{se0-f} \ee
 Thus, deviation of $s_{\varepsilon0}(T)$ from the ideal signal $s(T)$ is of the order of 
 $(\xi \sqrt n)^2$.

 Next, we calculate the second term on the right hand side of Eq.~(\ref{se}),
 which is the contribution of those multiplication terms that include
 only one $(\varepsilon_y \sigma_y)$ term in each of them.
 Noticing that, up to the second-order contribution of $\varepsilon_y$, 
 $\varepsilon_x $ in this second term can be taken as 1, 
 we have the following expression for it, 
 \bey
 s_{\varepsilon1}(T)=\mathfrak{Re}\la\downarrow|{R_{\varepsilon1}^\dagger}{\sigma_x}
 R_{\varepsilon0}+{R_{\varepsilon0}^\dagger}
 {\sigma_x}R_{\varepsilon1}|\uparrow\ra , \ \
 \label{s1} \eey
 where 
\bey
R_{\varepsilon1}=U(T,t_n)\varepsilon_{y,n}\sigma_yU(t_n,
t_{n-1})\sigma_x \cdots \nonumber
U(t_2,t_1)\sigma_xU(t_1,0)\nonumber
\\+U(T,t_n)\sigma_xU(t_n,t_{n-1})
\varepsilon_{y,n-1}\sigma_y \cdots \nonumber
U(t_2,t_1)\sigma_xU(t_1,0)\nonumber
\\+ \cdots \hspace{7cm}  \nonumber
\\ +U(T,t_n)\sigma_xU(t_n,t_{n-1})
\sigma_x \cdots U(t_2,t_1)\varepsilon_{y,1}\sigma_yU(t_1,0). \ \ \
\eey
 Making use of the fact that $\sigma_y=i\sigma_x\sigma_z$ and $[\sigma_z,H]=0$,
 we find
\bey R_{\varepsilon1}=
% \sum_{k=1}^ni(-1)^{k+1}\varepsilon_{y,k}U(T,t_n)\sigma_xU
% (t_n,t_{n-1})\sigma_x \cdots \nonumber\\ U(t_2,t_1)\sigma_xU(t_1,0)\sigma_z\nonumber\\
 i \sum_{k=1}^n (-1)^{k+1}\varepsilon_{y,k}R_{\varepsilon0}\sigma_z.
 \label{R1} \eey
 Substituting Eq.~(\ref{Re0}) into Eq.~(\ref{R1}), then into Eq.~(\ref{s1}), we obtain
\bey
s_{\varepsilon1}(T)=\mathfrak{Re} \left [ 2i\sum_{k=1}^n(-1)^{k+1}\varepsilon_{y,k}
\la\downarrow|{R_{\varepsilon0}^\dagger} {\sigma_x} R_{\varepsilon0}|\uparrow\ra \right ] .
\eey
 Typically, one has the following estimate,
\bey
 \left | \sum_{k=1}^n(-1)^{k+1}\varepsilon_{y,k} \right | \sim \xi {\sqrt{n}},
\eey
 hence,
\bey
s_{\varepsilon1}(T)\sim \pm 2\xi {\sqrt{n}} \mathfrak{Re} \left [ i\la\downarrow|
{R_{\varepsilon0}^\dagger}{\sigma_x} R_{\varepsilon0}|\uparrow\ra \right ] . 
\eey
 Making use of Eq.~(\ref{Re0}), after simple algebra, it is found that
\bey
s_{\varepsilon1}(T)\sim \pm 2\xi {\sqrt{n}}  q(T),
% \prod_{k=1}^n(1-\varepsilon_{y,k}^2), \
%\nonumber\\
%&=&-2\xi {\sqrt{n}} \mathfrak{Im}[\la\downarrow|{R_{\varepsilon0}^\dagger}{\sigma_x}
%R_{\varepsilon0}|\uparrow\ra]
%\nonumber
%\nonumber
% \\ &=&-2\xi {\sqrt{n}}\prod_{k=1}^n(1-\varepsilon_{y,k}^2) \mathcal{O}({T}^{2n+2})
 \label{se1-f} \eey
 where
 \be q(T) = \mathfrak{Im}\la\downarrow|{R^\dagger}{\sigma_x}R|\uparrow\ra .
 \label{qt}  \ee
 Therefore, $s_{\varepsilon1}(T)$ gives a first order correction ($\sim \xi \sqrt n$) 
 to the ideal signal.

 The first order correction $s_{\varepsilon1}(T)$ also depends on the quantity $q(T)$. 
 To give an estimation to $q(T)$, we note that 
 $s(T) + i q(T) = \la\downarrow|{R^\dagger}{\sigma_x}R|\uparrow\ra $,
 [see Eqs.~(\ref{sT}) and (\ref{qt})].
 Hence,
 \be |q(T)| = \sqrt{|\la\downarrow|{R^\dagger}{\sigma_x}R|\uparrow\ra|^2 -s^2(T)}
 \le \sqrt{1-s^2(T)}, \label{q-es}
 \ee
 where we have used the fact that
 $|\la\downarrow|{R^\dagger}{\sigma_x}R|\uparrow\ra|\leq1$.

 Finally, we discuss the third term on the right hand side of Eq.~(\ref{se}),
 which includes all multiplication terms that have only two $(\varepsilon_y \sigma_y)$ terms,
\bey
s_{\varepsilon 2}(T)=\mathfrak{Re}\la\downarrow|{R_{\varepsilon1}^\dagger}{\sigma_x}
R_{\varepsilon1}+{R_{\varepsilon0}^\dagger}{\sigma_x}R_{\varepsilon2}
+{R_{\varepsilon2}^\dagger}{\sigma_x}R_{\varepsilon0}|\uparrow\ra , \ \
  \label{se2} \eey
where
\bey
R_{\varepsilon2}&=&U(T,t_n)\varepsilon_{y,n}\sigma_yU(t_n,t_{n-1})\varepsilon_{y,n-1}
\sigma_y\nonumber\\
&&U(t_n,t_{n-1})\sigma_x \cdots U(t_2,t_1)\sigma_xU(t_1,0)
\nonumber
\\
&+&U(T,t_n)\varepsilon_{y,n}\sigma_yU(t_n,t_{n-1})\sigma_x\nonumber\\
&&U(t_n,t_{n-1})\varepsilon_{y,n-2}\sigma_y \cdots U(t_2,t_1)\sigma_xU(t_1,0)
\nonumber
\\
&+& \cdots +U(T,t_n)\sigma_xU(t_n,t_{n-1})\sigma_x\nonumber\\
&& \cdots U(t_3,t_2)\varepsilon_{y,2}\sigma_yU(t_2,t_1)\varepsilon_{y,1}\sigma_yU(t_1,0).
\eey
 Following a procedure similar to that for $R_{\varepsilon1}$, we find
\bey
R_{\varepsilon2} =
% \sum_{j,k=1,j<k}^n
 \sum_{k=1}^n \sum_{j(<k)}
 (-1)^{j+k+1} \varepsilon_{y,j}\varepsilon_{y,k}R_{\varepsilon0}
\sim \pm \frac{1}{2} {\xi^2}{n} R_{\varepsilon0}. \label{Re-2}
\eey

 Substituting the above obtained expressions of $R_{\varepsilon 0}$, $R_{\varepsilon1}$, 
 and $R_{\varepsilon2}$ into the expression of $s_{\varepsilon 2}(T)$ in Eq.~(\ref{se2}),
 after some derivation, we obtain an expression for $s_{\varepsilon 2}(T)$.
 Then, making use of results obtained above in Eqs.(\ref{se0-f}) and (\ref{se1-f}),
 finally, we find
 \bey
 s_\varepsilon(T) - s(T) \sim \pm 2q(T) \xi {\sqrt{n}} +  C_2{\xi^2}{n} + O(\xi^3) , \label{s-ep}
 \eey
 where $|C_2|$ is of the order of 1.

 Of particular interest is the case for a DD with good performance, i.e., with $s(T) \sim 1$.
 In this case, the inequality  (\ref{q-es}) shows that $q(t)$ is small.
 In particular, in the case that $|q(t)| \ll \xi \sqrt{n}$, 
 the first order term on the right hand side of
 Eq.~(\ref{s-ep}) can be neglected and we have
 \bey
 s_\varepsilon(T) - s(T) \sim  C_2{\xi^2}{n} + O(\xi^3) , \label{s-ep-1}
 \eey
 scaling as $\xi^2n$.
 For example, for a UDD with $s(T)=1-{O}({T}^{2n+2})$,
 $q(T)$ is of the order of ${O}({T}^{2n+2})$ or less, hence, for a fixed $\xi$, 
 $|q(T)| \ll \xi \sqrt{n}$ for a sufficiently large $n$.

 In concluding this section, we remark that it is straightforward to generalize the 
 above discussions to the case of pulses with more generic random errors in their direction.
 In fact, in this generic case,  the time-evolution operator can be obtained by replacing $\sigma_x$
 in Eq.~(\ref{R0}) by
 \be \sigma_\varepsilon = \varepsilon_x \sigma_x+\varepsilon_y\sigma_y
 +\varepsilon_z\sigma_z, \label{sig-e-g}
 \ee
 where $\varepsilon_x =\sqrt{1-\varepsilon_y^2-\varepsilon_z^2}$ and
 both $\varepsilon_y$ and $\varepsilon_z$ are small random numbers with Gaussian distribution.
 We have found results that are qualitatively the same as those discussed above 
 for the case of $\sigma_\varepsilon$ in Eq.~(\ref{sig-e}).
 In particular, we have found similar estimations as those given in the relations
 (\ref{s-ep}) and (\ref{s-ep-1}).

 \section{Numerical simulations}
 \label{sect-num}

 In this section, we discuss numerical simulations we have performed for two purposes.
 One is to check analytical predictions given in the previous section,
 the other is to study the influence of kicks in the environment in the efficiency of UDD.
 In fact, since instant kicks in the environment may have non-negligible high-frequency components, 
 the performance of UDD for such an environment may be not so good as that for an environment
 with a sharp high-frequency cutoff.

%% Figure 1
\begin{figure}
\includegraphics[width=8cm]{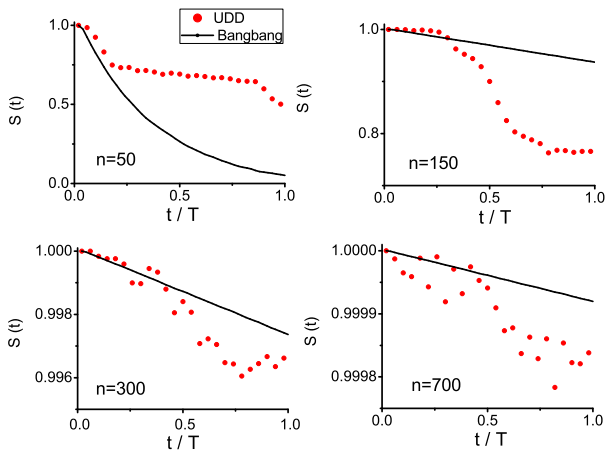}
 \caption{(Color online)
 Variation of the signal $s(t)$ in Eq.~(\ref{s-t}) with time  
 for ideal UDD and ideal periodic bang-bang  control.
 Both UDD and periodic bang-bang  control have better performance for larger values of $n$,
 the number of controlling pulses.
 Results of UDD (circles) are better than those of the periodic bang-bang  control
 (solid curve) for $n=50$. 
 But, for larger values of $n$, the periodic bang-bang  control outperforms UDD.
 Parameters are
 $\omega_A=1.5\times10^3, \lambda=10^3, g=100, N=2^{12}$ and $T=50T_0$,
 and the kicked rotator is in the chaotic region with $K=10^{3}$.
 The same parameters are also used in the following figures.
 }
  \label{fig-UDD-bb}
 \end{figure}

 \subsection{The model}

 We consider a model, in which there is a qubit $S$ of interest and an environment $\E$ 
 that is composed of a second qubit $A$ and a quantum kicked rotator denoted by $B$.
 The qubit $S$ has interaction with $A$ only, while $A$ interacts with both $S$ and the kicked
 rotator $B$.
 The Hamiltonian is written as
 \begin{eqnarray}
  H=H_{S}+H_{A}+H_{B}+H_{SA}+H_{AB} , \label{H1}
 \end{eqnarray}
 where the self-Hamiltonians are
 $H_{S}=0$,  $H_{A}$=$\omega_A\sigma_{x}^{A}$, and
 \be
   H_{B}=\frac{p^2}{2}+k\cos\theta \sum_{j} \delta(t-jT_0), \label{H-kr}
 \ee
 with $T_0$ the period of kicking.
 The interaction Hamiltonians are
 $H_{SA} = g \sigma^{S}_{z}\otimes\sigma^{A}_{z}$ and
 \begin{eqnarray}
 H_{AB}=\lambda\sigma^{A}_{z}\cos\theta\sum_{j}\delta(t-jT_0).
 \label{H-AB} \end{eqnarray}
 Here, for clearness we write explicitly the superscript $S$ in the Pauli operator for the qubit $S$.

 This model has been studied in Ref.~\cite{RCB06},
% with both qubits coupled to the kicked rotator,
 showing that the kicked rotator,
 though a single-particle dynamical system, may simulate a pure-dephasing
 many-body bath (Caldeira-Leggett model \cite{CL83}), as well as some non-Markovian environments.
 Here, we are interested in the chaotic region of the kicked rotator, to simulate
 some random properties of the environment.

%% Figure 4
\begin{figure}
\includegraphics[width=\columnwidth]{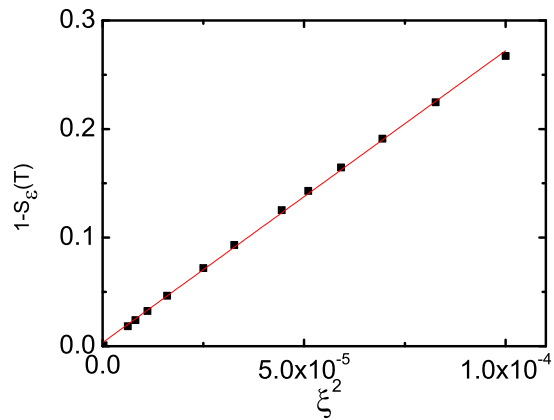}
 \caption{(Color online) Dependence of $1-s_\varepsilon(T)$ (solid squares) 
 on $\xi^2$ for a UDD sequence, 
 with $n=500$ and $\xi \sqrt n \ll 1$. 
 The solid curve is a straight line, indicating a linear dependence,
 in agreement with the prediction of Eq.~(\ref{s-ep-1}). 
 For the corresponding ideal UDD,  $1-s(T)\sim10^{-4}$. 
 Since $1-s_\varepsilon(T) \gg 1-s(T)$, the influence of the pulse errors is profound. 
 }
  \label{fig-1-s-small}
 \end{figure}

 The time evolution for one period $T_0$ is given by the unitary operator
 \bey
 \hat{U}_{T_0} = e^{-i(\omega_{A}T_0 {\sigma}^{A}_x+
 T_0g {\sigma}^{S}_{z}\otimes {\sigma}^{A}_{z})}
 e^{-iT_0\frac{ {p}^2}{2}}e^{-i(k+\lambda {\sigma}^{A}_{z})\cos {\theta}},
% e^{-i\lambda {\sigma}^{A}_{z}\cos {\theta}}, \
 \eey
 where $\hbar$ has been set unit.
 An effective Planck constant can be introduced, $\hbar_{\rm eff}=T_0=2\pi/N$,
 where $N$ is the dimension of the Hilbert space of the kicked rotator.
 The classical limit is obtained by letting $T_0\to 0$ and $k\to\infty$ while
 keeping  $K=kT_0$ fixed,
 and the classical counterpart is defined on a torus $[0,2\pi ) \otimes [0,2\pi )$.
 The kicked rotator has a chaotic motion for $K$ larger than 6 or so.
% As a result of quantization on torus, the upper frequency is limited by the upper bound of
% the momentum $p$, namely, $2\pi$.

%% Figure 5
\begin{figure}
\includegraphics[width=\columnwidth]{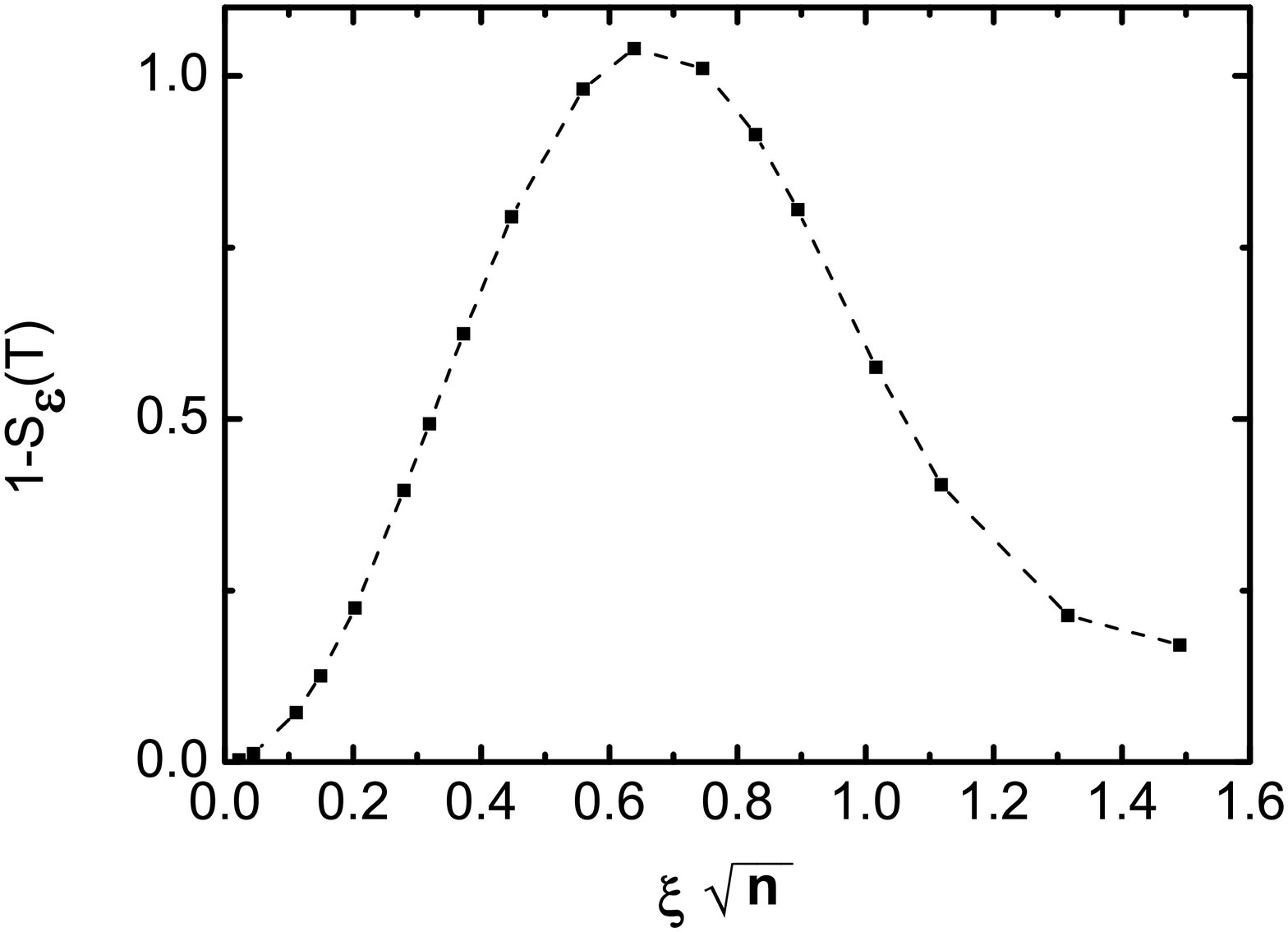}
 \caption{ Variation of $1-s_\varepsilon(T)$ with $\xi\sqrt{n}$ for a UDD sequence with
 $n=500$.
 }
  \label{fig-1-s-large}
 \end{figure}

 \subsection{Numerical results}

 Let us first discuss the performance of ideal UDD and ideal periodic bang-bang  control in this model. 
 We are interested in the case that $T \gg T_0$.
 In this case, due to the time-dependency of the Hamiltonian of the environment,
 Uhrig's strategy for deriving $1-s(T) \sim {O}({T}^{2n+2})$ is not applicable.
 It is of interest to know whether UDD is optimal or not in this case.

 In numerical simulation, we calculated $s(t)$ defined by
 \bey
 s(t)=\mathfrak{Re}\la\downarrow|{R^\dagger(t)}{\sigma_x}R(t)|\uparrow\ra , \label{s-t}
 \eey
 where $R(t)$ is obtained by truncation of the time evolution operator $R$ in Eq.~(\ref{R0})
 at an intermediate time $t \le T$. 
 Figure \ref{fig-UDD-bb} shows that both UDD and periodic bang-bang  control have better 
 performance with increasing number $n$ of the controlling pulses. 
 For $n=50$, UDD outperforms the periodic bang-bang  control in the whole time region $(0,T]$.
 However, with increasing $n$, results of the periodic bang-bang  control become better than those of UDD
 for $t>T/2$.

 Next, we check analytical predictions given above 
 for the impact of random errors in the direction of the controlling pulses,
 in particular, the behavior of $s_\varepsilon(T)$ in Eq.~(\ref{s-ep-1}). 
 We use the general form of $\sigma_\varepsilon$ in Eq.~(\ref{sig-e-g}), with the same dispersion
 $\xi$ for $\varepsilon_y$ and $\varepsilon_z$.
 An example is given in Fig.\ref{fig-1-s-small}, showing linear
 dependence of $1-s_\varepsilon(T)$ on $\xi^2$.
 To check details of agreement with analytical predictions,  we have numerically computed  
 the corresponding ideal UDD and found that it has a good performance with $1-s(T)\sim10^{-4}$. 
 This gives $|q(T)| \lesssim 10^{-2}$ [see the estimate in the inequality (\ref{q-es})]. 
 Hence, $|q(T)| \ll \xi \sqrt{n}\sim10^{-1}$ for solid squares shown in the figure,
 as a result, Eq.~(\ref{s-ep-1}) should hold with $s(T)-s_\varepsilon(T) \simeq 1-s_\varepsilon(T)$. 
 Thus, results in Fig.\ref{fig-1-s-small} indeed confirm the prediction of Eq.~(\ref{s-ep-1}).

%% Figure 2
\begin{figure}
\includegraphics[width=\columnwidth]{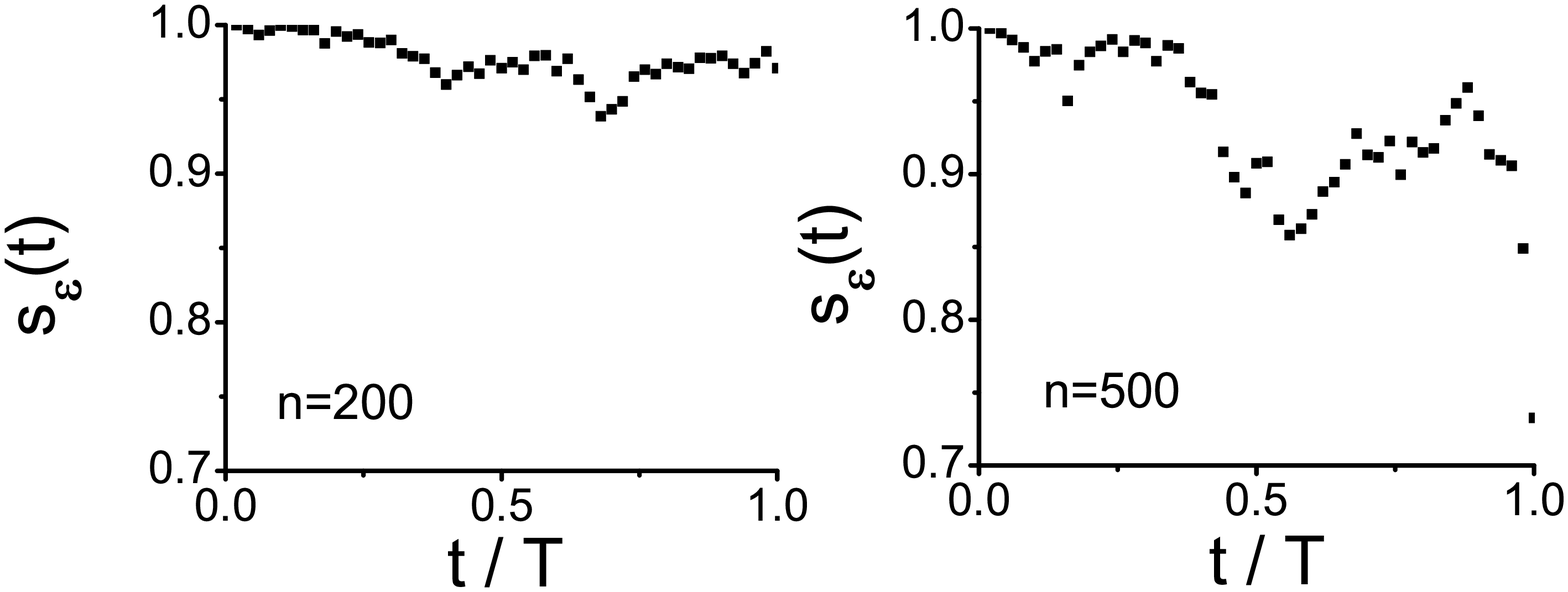} %\\[-2.5cm]
 \vspace{-3.2cm}
 \caption{ 
 Variation of the signal $s_\varepsilon(t)$ with time for a UDD with non-ideal pulses,
 $\xi = 10^{-2}$.
 The UDD with $n=500$ (right panel) gives results worse than those of the UDD with $n=200$
 (left panel). 
 }
 \label{fig-dev}
 \end{figure}

 We have also numerically studied the case of large $\xi \sqrt{n}$.
 In this case, higher order terms of $\xi$  Eq.~(\ref{s-ep}) should be considered.
 Anyway, this expression of $s_\varepsilon(T)$ suggests that
 $1-s_\varepsilon(T)$ may be large for $\xi^2 n \sim 1$.
 Indeed, numerical results support this expectation, as shown in Fig.~\ref{fig-1-s-large}.

 Figure \ref{fig-1-s-small} shows that for a fixed $n$, 
 the influence of pulse errors may become large with increasing $\xi$, 
 such that $1-s_\varepsilon(T) \gg 1-s(T)$. 
 In Fig.\ref{fig-dev}, we show that for a fixed error dispersion $\xi$, 
 deviation of $s_\varepsilon(t)$ from 1 enlarges when $n$ is increased,
 where $s_\varepsilon(t)$ is defined by Eq.~(\ref{s-t}) with $R(t)$ replaced by
 the corresponding $R_\varepsilon(t)$.

 \section{Conclusions and discussions}
 \label{sect-conc}

 In this paper, we have analytically studied the efficiency of a generic DD with a single-axis control,
 when the controlling pulses are subject to small random errors in their direction.
 We have derived expressions for the influence of the pulse errors up to the second-order term.
 When the ideal DD has a sufficiently good performance, the influence has
 a negligible first-order effect; in this sense, good DD are relatively robust.

 We have tested the above analytical predictions numerically and shown that accumulation of 
 small pulse errors may have significant influence in the efficiency of DD. 
 For an environment with kicks applied on some part of the environment, 
 it has been found that the periodic bang-bang  control may outperform UDD.

 A natural question would concern the possibility of having negligible first-order effect of 
 pulse errors in more generic situations, e.g., in the case of more than one layers of 
 controlling pulses with different single-axis control in different layers. 
 For pulse errors appearing only at certain fixed layer with single-axis control, 
 results of this paper may be generalizable.
 However, the more generic situation with pulse errors in different layers, as well as the 
 case with finite width of the pulses, are much more 
 complex and further investigation is needed before a definite conclusion can be drawn.

 ACKNOWLEDGMENTS.
 The authors are grateful to Jiangbin Gong for valuable discussions and suggestions.
 This work is partly supported by the Natural Science Foundation of China under Grant Nos.~10775123
 and 10975123 and the National Fundamental Research Programme of China
 Grant No.2007CB925200.

 \end{document}